%% file: main.tex
\documentclass[]{spie}  %

\usepackage{amsmath,amsfonts,amssymb}
\usepackage{graphicx}
\usepackage[colorlinks=true, allcolors=blue]{hyperref}
\input{packages}

\input{macro}

\title{Toward Intelligent Electronic-Photonic Design Automation for Large-Scale Photonic Integrated Circuits: from Device Inverse Design to Physical Layout Generation}

\author[a]{Hongjian Zhou\textsuperscript{*}}
\author[a]{Pingchuan Ma\textsuperscript{*}}
\author[a]{Jiaqi Gu}
\affil[a]{School of Electrical, Computer and Energy Engineering, Arizona State University}

\authorinfo{Further author information: (Send correspondence to Jiaqi Gu)\\Jiaqi Gu: E-mail: jiaqigu@asu.edu}

\begin{document} 
\maketitle

\input{doc/0_abstract}


\input{doc/1_intro}

\input{doc/2_flow}

\input{doc/3_EPDA}

\input{doc/4_Direction}

\input{main.bbl}

\end{document}

%% file: packages.tex
\usepackage[utf8]{inputenc} %
\usepackage[T1]{fontenc}    %
\usepackage{pifont}
\usepackage{xcolor}         %
\usepackage{booktabs}
\usepackage{hyperref}
\hypersetup{colorlinks,
    linkcolor=red,citecolor=citecolor,urlcolor=blue}
\usepackage{algorithm}
\usepackage{graphicx}
\usepackage{threeparttable}
\usepackage{multirow}
\usepackage{amsmath}
\usepackage{bm}
\usepackage{bbm}
\usepackage{amsfonts}
\usepackage[]{algpseudocode}
\usepackage{enumitem} %
\usepackage[subrefformat=parens,labelformat=parens]{subfig}
\usepackage{colortbl}
\usepackage{geometry}

\usepackage{wrapfig}
\usepackage[algo2e, ruled, vlined, linesnumbered, noend]{algorithm2e}

%% file: macro.tex
\definecolor{citecolor}{RGB}{34,139,34}
\definecolor{mydarkblue}{rgb}{0,0.08,1}
\definecolor{mydarkgreen}{rgb}{0.02,0.6,0.02}
\definecolor{mydarkred}{rgb}{0.8,0.02,0.02}
\definecolor{mydarkorange}{rgb}{0.40,0.2,0.02}
\definecolor{mypurple}{RGB}{111,0,255}
\definecolor{myred}{rgb}{1.0,0.0,0.0}
\definecolor{mygold}{rgb}{0.75,0.6,0.12}
\definecolor{myblue}{rgb}{0,0.2,0.8}
\definecolor{mydarkgray}{rgb}{0.,0.2,0.2}

\definecolor{lightred}{RGB}{255,235,235}
\definecolor{lightgreen}{RGB}{235,255,235}
\definecolor{lightblue}{RGB}{235,235,255}
\definecolor{lightcyan}{RGB}{235,255,255}
\definecolor{lightmagenta}{RGB}{255,235,255}
\definecolor{lightyellow}{RGB}{255,255,235}

\definecolor{qxkcolor}{RGB}{215,235,255}
\definecolor{softmaxcolor}{RGB}{230,235,255}
\definecolor{probxvcolor}{RGB}{255,255,235}

\definecolor{topkcolor}{RGB}{255,235,235}
\definecolor{zecolor}{RGB}{255,255,235}
\definecolor{dynacolor}{RGB}{235,255,255}

\definecolor{reviewcolor}{RGB}{0,0,200}

\algdef{SE}[DOWHILE]{Do}{doWhile}{\algorithmicdo}[1]{\algorithmicwhile\ #1}%

\newcommand{\squishlist}{
 \begin{list}{$\bullet$}
  { \setlength{\itemsep}{0pt}
     \setlength{\parsep}{3pt}
     \setlength{\topsep}{3pt}
     \setlength{\partopsep}{0pt}
     \setlength{\leftmargin}{1.5em}
     \setlength{\labelwidth}{1em}
     \setlength{\labelsep}{0.5em} } }

\newcommand{\squishend}{
  \end{list}  }

%% file: doc/0_abstract.tex
\section{abstract}
\label{sec:Abstract}

Photonic Integrated Circuits (PICs) offer tremendous advantages in bandwidth, parallelism, and energy efficiency, making them essential for emerging applications in artificial intelligence (AI), high-performance computing (HPC), sensing, and communications. 
However, the design of modern PICs, which now integrate hundreds to thousands of components, remains largely manual, resulting in inefficiency, poor scalability, and susceptibility to errors. 
To address these challenges, we propose PoLaRIS, a comprehensive Intelligent Electronic-Photonic Design Automation (EPDA) framework that spans both device-level synthesis and system-level physical layout. PoLaRIS combines a robust, fabrication-aware inverse design engine with a routing-informed placement and curvy-aware detailed router, enabling the automated generation of design rule violation (DRV)-free and performance-optimized layouts. 
By unifying physics-driven optimization with machine learning and domain-specific algorithms, PoLaRIS significantly accelerates PIC development, lowers design barriers, and lays the groundwork for scalable photonic system design automation.

%% file: doc/1_intro.tex
\section{INTRODUCTION}
\label{sec:INTRODUCTION} 
Driven by the demands of artificial intelligence (AI)~\cite{NP_Nature_ahmed}, high-performance computing (HPC)~\cite{NP_Light_Zhou}, and optical interconnects~\cite{NP_DATE2020_popstar}, photonic integrated circuits (PICs) are undergoing a rapid increase in scale and complexity, with modern designs integrating thousands of components on a single chip. 
This explosive growth has rendered traditional manual design methodologies~\cite{NP_book_chrostowski} a critical bottleneck. The manual layout process is not only time-consuming and error-prone but is fundamentally unscalable for contemporary PIC systems, making robust automation tools essential for future progress.
To address this need, we focus on two major fronts in photonic design automation: circuit-level physical layout, encompassing PIC placement and routing (PnR), and device-level synthesis, largely driven by inverse design. 
PnR tools aim to solve the physical arrangement and interconnection of components on a chip, while inverse design seeks to optimize the topology of individual photonic components from functional specifications to achieve ultra-compact, high-performance structures.

Existing PnR approaches often directly adapt electrical-circuit algorithms to photonics, mainly addressing high-level constraints such as minimizing waveguide crossings and insertion loss. 
However, they \emph{overlook critical photonics-specific requirements and physical implementation challenges}. 
Meanwhile, inverse design techniques are believed to be capable of discovering compact, high-performance devices.
However, inverse-designed structures often suffer from post-fabrication performance degradation, process variations, and the prohibitive computational cost of repeated simulations, which ultimately hinders broad adoption.

To overcome these hurdles, we propose PoLaRIS (\underline{P}h\underline{o}tonic \underline{La}yout, \underline{R}outing \& \underline{I}nver\underline{s}e Device Design), comprising two main components. 
(1) \textbf{PoLaRIS-invdes} is an advanced adjoint-based inverse design toolkit that abstracts complex physics yet supports flexible optimization steps. By integrating pre-trained AI models and fabrication-aware variation modeling, it ensures real-world applicability and speeds up the optimization process.
 (2) \textbf{PoLaRIS-PnR} is an automated placement and routing tool featuring a differentiable photonic placer and a "curvy-aware" routing engine with adaptive crossing insertion, effectively handling waveguide bends and crossings while minimizing insertion loss.
 \input{figtex/fig_design_flow}

%% file: figtex/fig_design_flow.tex
\begin{figure}
    \centering
    \includegraphics[width=\columnwidth]{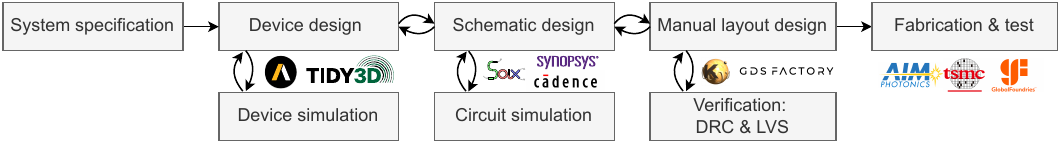}
    \vspace{-5pt}
    \caption{Simplified PIC design flow and example automation toolkits and foundry vendors.
    }
    \label{fig:design_flow}
     \vspace{-5pt}
\end{figure}

%% file: doc/2_flow.tex
\section{Overview of Traditional Flow}
\label{sec:flow}

In this section, we first outline the current PIC design flow with a focus on the physical implementation stage, including both device-level design and full-layout generation. 
We highlight key challenges in this process and review existing methods aimed at addressing them.

The conventional PIC design flow typically follows a \emph{sequential, manual} pipeline~\cite{NP_book_chrostowski} involving device design, schematic construction, layout drafting, and verification as shown in Fig.~\ref{fig:design_flow}. 
Custom components such as modulators, filters, and multiplexers are often designed manually, then connected into systems via schematic capture tools, followed by manual or rule-based layout and final design rule checking.
While this flow is currently manageable by experienced designers, it relies heavily on hierarchical design methodologies. 
Hierarchy enables parallelism and specialization across teams, such as PDK development, functional block design, and top-level layout, but it also introduces \emph{inefficiencies}. 
Schematic design depends on accurate yet time-consuming device models derived from complex electromagnetic (EM) simulations. 
Layout design is also slow and often requires iterative refinement, with back-annotation cycles to update the schematic for accurate simulation.
Each stage also demands \textbf{deep domain expertise}, creating barriers for new designers and interdisciplinary researchers. 
\textbf{This fragmentation and high cross-domain barrier limits accessibility and slows innovation}.

To address these challenges, there is an urgent need for a \emph{re-defined design automation toolflow} across the PIC design stack. 
An ideal workflow would allow designers to create high-performance optical components and circuits with fast, accurate simulation and automated layout generation, freeing them to focus on system-level topology and architecture.
Such automation would reduce iteration cycles, enable broader design space exploration, and accelerate time-to-market, while also improving design quality through more frequent and intelligently guided refinement. 
By lowering the entry barrier, it would also broaden accessibility for students, researchers, and engineers, ultimately advancing innovation in integrated photonics.
This work aims to \textbf{advance the automation on both device design and layout generation processes}. 
In the following, we highlight the key challenges of each and review existing methodologies.

\subsection{PIC Design Automation Challenges}

\subsubsection{Device Inverse Design Challenges} 
In device design, the engineer develops and optimizes a component for a specific function, then hands off its abstracted model (e.g., S-parameters) to the circuit designer, who links multiple such devices together to form a complete circuit schematic. 
When designing the devices for PICs, the goal is to optimize the device structures to fulfill a specific optical function and figure-of-merits (FoMs).

Traditional photonic device design typically relies on \emph{extensive physics-based prior knowledge and iterative trial-and-error} approaches, placing significant demands on designers. 
Such methods tend to be inefficient, requiring substantial manual tuning and often resulting in relatively large device footprints due to limited exploration of intuitive design spaces. 
In contrast, \textbf{inverse design}~\cite{NP_minkov2020inverse} formulates device design as an optimization problem guided by clearly defined objectives and constraints, substantially reducing the reliance on detailed physics expertise or heuristics. 
By enabling exploration of high-dimensional, non-intuitive design spaces, inverse design allows for more compact devices than traditional methods typically achieve. 
Additionally, the adjoint method~\cite{NP_khoram2020controlling} further improves efficiency by analytically computing gradients using only two forward electromagnetic simulations, greatly accelerating the optimization process with higher optimality over traditional manual search or parameter sweeping.
\input{figtex/fig_invDes_challenge}
Although \underline{adjoint-based inverse design} has shown great promise in efficiently optimizing photonic devices, it faces several \underline{intrinsic optimization challenges}, as shown in Fig.~\ref{fig:invdes_challenge}. 
\ding{202}~First, adjoint optimization is typically formulated as a high-dimensional, discrete optimization problem characterized by a complex and non-convex loss landscape. This complexity results in numerous local optima, making the optimization highly sensitive to initial conditions and prone to convergence toward suboptimal solutions. 
\ding{203}~Second, the objective function is typically sparsely defined, relying primarily on performance metrics evaluated at device output ports, which leads to limited gradient information and vanishing gradients, resulting in unstable optimization dynamics.
\ding{204}~Additionally, each iteration of adjoint optimization requires multiple computationally expensive numerical simulations to calculate gradients, making the optimization process extremely time-consuming and poorly scalable to larger, more complex design problems.
Recent advances in \emph{scientific machine learning} have inspired several neural-network (NN)-based surrogate models~\cite{NP_li2020fourier, NP_tran2021factorized, NP_zhang2024sinenet} to replace traditional numerical solvers within the inverse design loop, significantly accelerating optimization. 
However, existing neural networks are typically \textbf{trained using datasets and objectives that do not align well with their intended optimization tasks}\cite{gu2022NeurOLight, zhu2024Pace, kossaifi2024neural, kovachki2021neural, zhang2024sinenet, pcma2024pic2osim}. 
This mismatch leads to surrogate models that produce \emph{visually plausible results yet struggle to generalize} across diverse design spaces and fail to provide accurate gradients required for reliable inverse design.

While adjoint-based inverse design can yield numerically plausible results, these optimized structures often experience significant performance degradation when fabricated and deployed in real environments. 
A key reason for this gap is that inverse-designed structures frequently contain complicated features at very small scales. 
Such \textbf{tiny features are highly sensitive to fabrication imperfections} and typically cannot be accurately reproduced in practice. 
As the performance of inverse-designed devices critically depends on these fine-scale details, even minor discrepancies introduced during fabrication can severely degrade actual device functionality. 
Furthermore, practical deployment conditions are inherently \textbf{variable due to fabrication variations and environmental fluctuations}, as shown in Fig.~\ref{fig:var_source}, meaning the inverse design should, in fact, be formulated as a \emph{stochastic optimization problem to robustly account for these uncertainties}.
\underline{Previous approaches}~\cite{NP_chen2020design, NP_gershnabel2022reparameterization, hammond2022high, NP_khoram2020controlling} have attempted to address fabrication challenges in adjoint-based inverse design by incorporating heuristic methods that control minimum feature sizes (MFS). 
Techniques such as design reparameterization, curvature penalties, and low-pass filtering have been employed to eliminate tiny, non-manufacturable structures\cite{wang2019robust, wang2011robust, schevenels2011robust, gershnabel2022reparameterization, hammond2021photonic, chen2020design, mao2023multi}. 
However, these methods are only approximate representations of actual fabrication processes, as they do not accurately capture lithography and etching effects, leading to residual performance gaps after fabrication. 
Furthermore, earlier methods typically model fabrication variations by optimizing designs under simplified assumptions, such as uniform geometric dilation or erosion. 
While computationally inexpensive, this oversimplification fails to reflect real-world variations accurately, resulting in limited robustness improvements. 
Attempts to address robustness by \emph{exhaustive sampling of variation scenarios (so-called corners)} also incur \emph{prohibitive simulation costs}, scaling exponentially with the number of variation parameters, making these approaches impractical for realistic, complex device designs, as shown in Fig.~\ref{fig:adaptive_sampling}.
\input{figtex/fig_varSrc_adaptiveSample}

\subsubsection{Circuit Layout Generation Challenges}

After obtaining the netlist and component layouts, designers typically follow a schematic-driven layout (SDL) approach~\cite{chrostowski2016schematic}, manually placing components and connecting them with waveguides and electrical wires. 
This process must comply with complex photonic design rules, making it both time-consuming and error-prone.

Due to the single-layer silicon waveguide constraint in most silicon photonic PICs, all components and routes must share the same layer. 
This requires careful placement to reserve sufficient routing space while accounting for area-consuming features like bends and crossings. 
For example, single-mode waveguide bends generally require a minimum 5 $\mu\mathrm{m}$ radius to avoid loss, and crossings can occupy over $10\,\mu\mathrm{m} \times 10\,\mu\mathrm{m}$. 
Improper planning may lead to waveguide routing failure due to insufficient spacing.
The problem is further exacerbated by the need for precise port alignment and any-angle routing, waveguide port connections must be face-to-face, and crossings typically occur at 90°. 
Components with many ports increase bend and crossing density, further intensifying routing congestion and complexity.

In addition to photonic routing, electrical routing must also be addressed. 
Most PIC processes offer only \emph{two to three metal layers}, often \emph{without well-defined preferred routing directions} as in electronic ICs. 
This makes \textbf{via minimization} a key challenge, especially for DC signals, where input/output (IO) pads are placed along the chip periphery, resulting in long, hard-to-manage wires.
Moreover, for high-speed electrical signals, impedance matching becomes critical, adding further complexity and imposing stricter layout constraints.

To address the aforementioned challenges, \underline{current PIC layout workflows} often adopt the schematic-driven layout methodology, where waveguide \textbf{bends and crossings are treated as components and must be manually planned ahead during the schematic creation stage}. 
While tools like GDSFactory~\cite{GDSFactory} assist in automating this process, resolving layout conflicts and ensuring rule compliance still typically requires multiple rounds of manual iteration, making the process highly time-consuming.
In practice, script-based layout generation is also common. 
For a fixed topology, designers can rapidly generate layouts by predefining parameters such as spacing, bend radius, and alignment rules. 
However, this approach lacks scalability and generality.
Scripts are topology-specific and tightly coupled to a given PDK, making them difficult to reuse across designs or fabrication platforms.

Several previous works~\cite{NP_DAC2018_planaronoc, NP_DAC2009_ding, NP_ISPD2019_psion} have explored automated placement and routing for PICs, primarily targeting the reduction of wirelength, bends, and crossings to minimize insertion loss. 
However, many approaches overlook detailed photonic design rules and thus fail to guarantee fully legal layouts.
To bridge this gap, the design flow must advance beyond performance optimization to also enforce \textbf{complex photonic-specific constraints}, including automatic crossing insertion, minimum bend radii, port alignment, and manufacturability. 
A truly scalable and robust PIC layout automation framework should ensure flexibility, efficiency, and correct-by-construction layout generation across diverse topologies and fabrication processes.

%% file: figtex/fig_invDes_challenge.tex
\begin{figure}
    \centering
    \includegraphics[width=0.95\columnwidth]{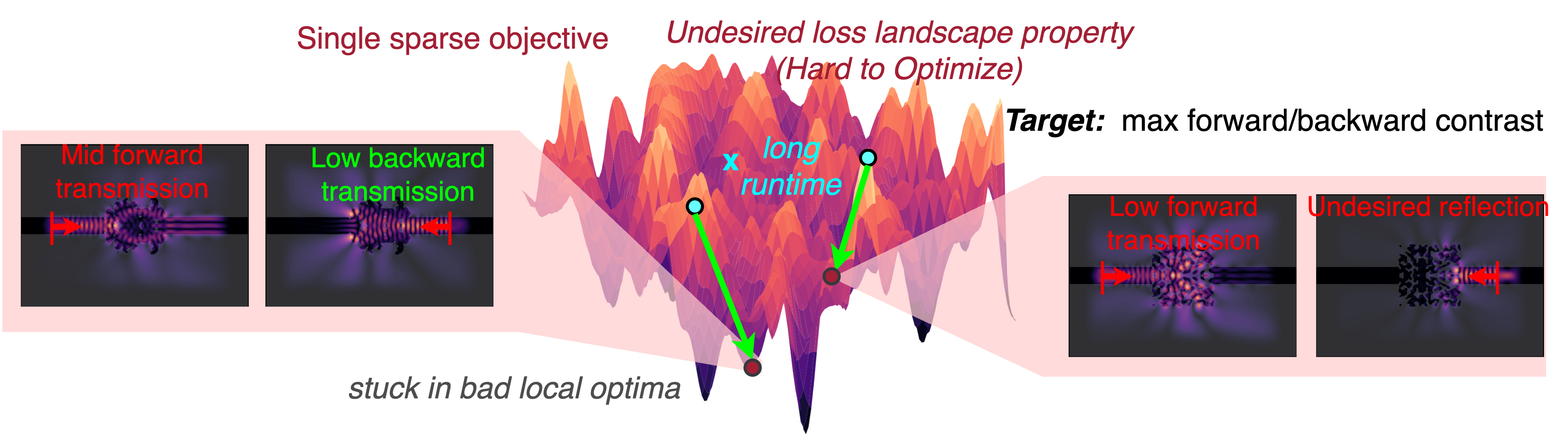}
    \vspace{-5pt}
    \caption{Inverse design, as a high-dimensional and non-convex optimization problem, faces three key challenges: a rugged and ill-conditioned loss landscape, unstable optimization dynamics, and prohibitively expensive, non-scalable computational costs.
    }
    \label{fig:invdes_challenge}
     \vspace{-5pt}
\end{figure}

%% file: figtex/fig_varSrc_adaptiveSample.tex
\begin{figure}
    \centering
    \subfloat[]{\includegraphics[width=0.47\columnwidth]{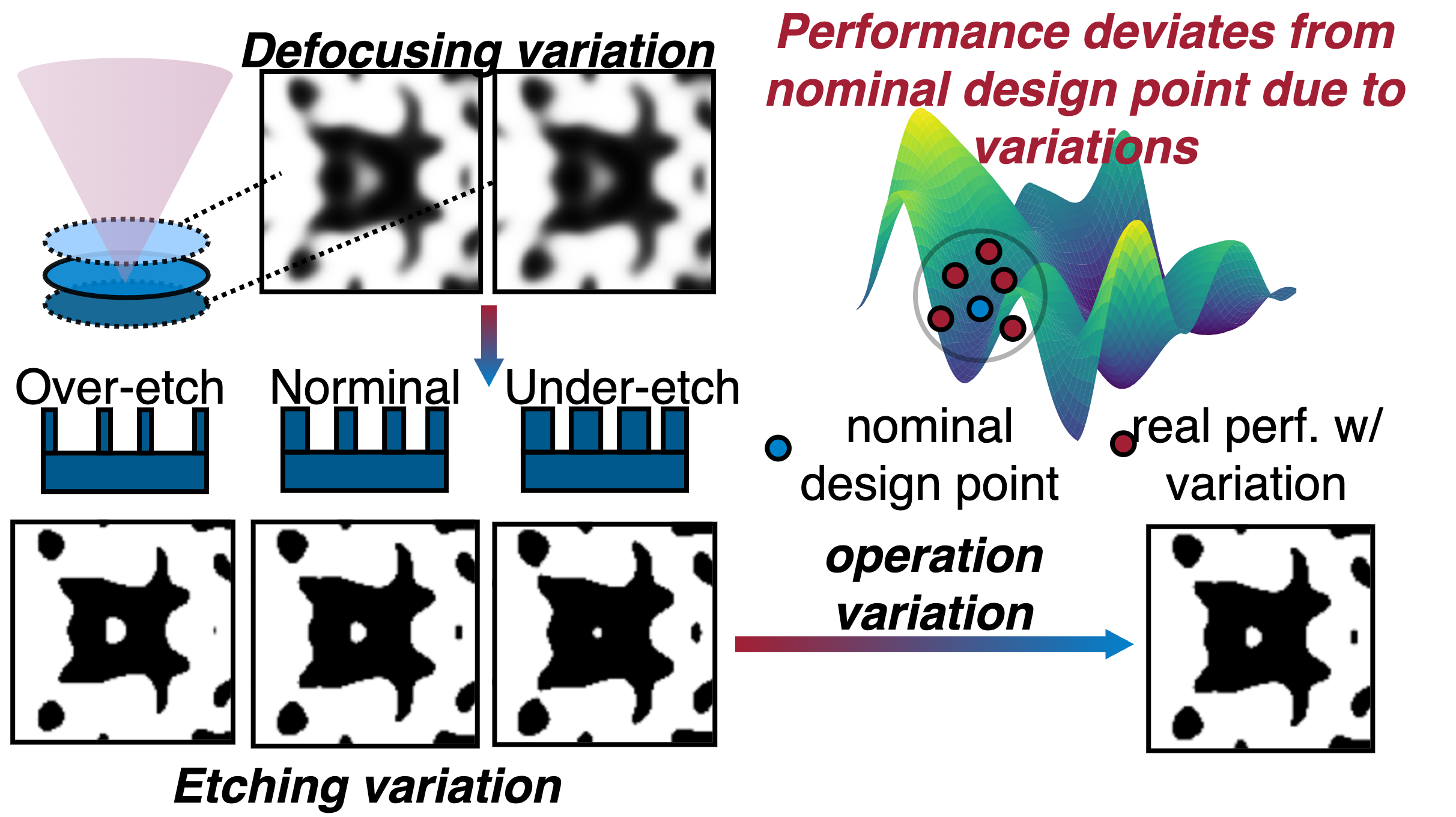}
    \label{fig:var_source}
    }
    \subfloat[]{\includegraphics[width=0.51\columnwidth]{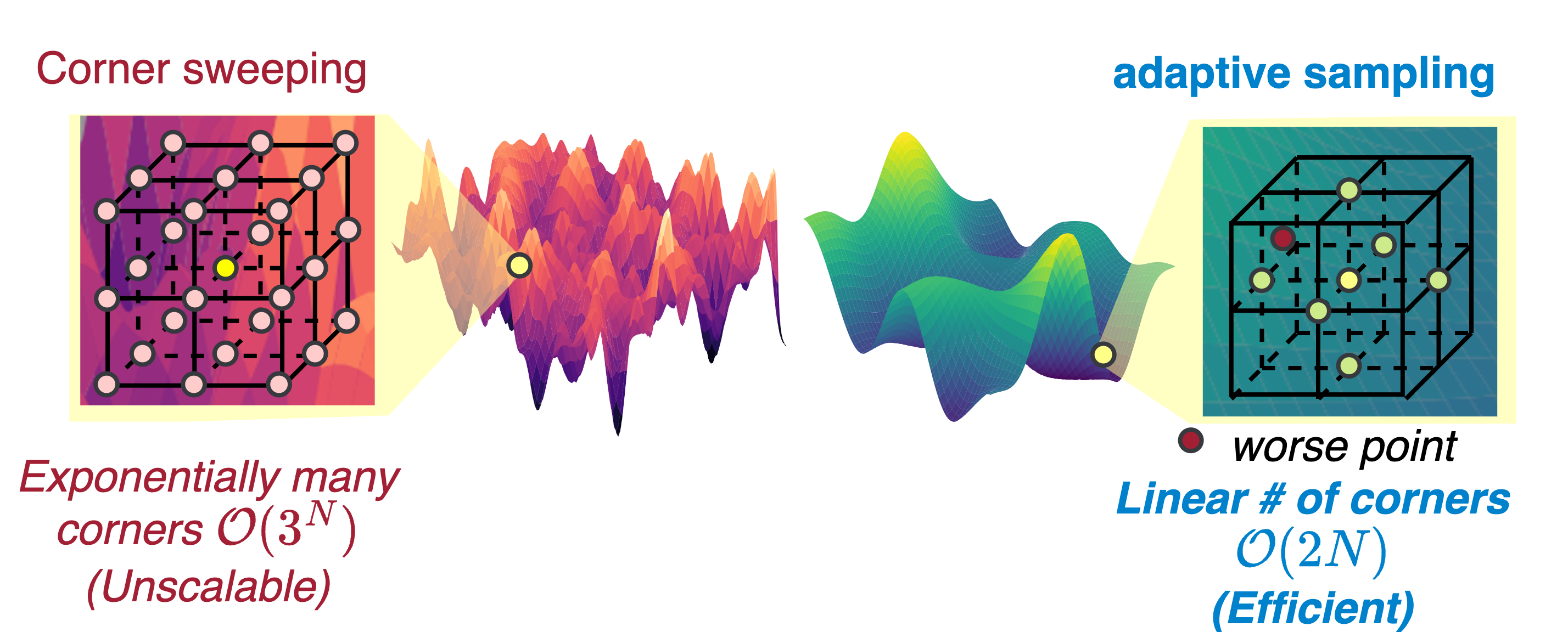}
    \label{fig:adaptive_sampling}
    }
    \caption{(a) From design pattern to device implementation, there are multiple sources of variations
    (b) Exhaustive sampling introduces unaffordable and unscalable optimization costs.
    }
    \label{fig:varSrc_adaptiveSam}
    \vspace{-10pt}
\end{figure}

%% file: doc/3_EPDA.tex
\section{Advanced EPDA design framework: PoLaRIS}
\label{sec:PoLaRIS}

In the preceding discussion, we highlight how device geometry directly affects each component’s functionality, footprint, insertion loss, and bandwidth. 
Meanwhile, the interconnects between components introduce additional insertion loss, crosstalk, and impact manufacturability. 
Although modern PIC design has borrowed key inspirations from electronic IC design flow, i.e., leveraging PDK libraries and hierarchical flows for better manageability, photonic circuits remain highly customized, making it difficult to directly apply existing EIC automation design algorithms to PIC workflows. 
To meet the demand of future large-scale PIC circuits, the existing design flow must be evolved. 
In this section, we propose PoLaRIS (Photonic Layout, Routing \& Inverse Device Design),
comprising two main components, spanning device‐level and circuit‐level generation to dramatically accelerate PIC design.

\input{doc/3_1_InvDesign}

\input{doc/3_2_PhyDesign}

%% file: doc/3_1_InvDesign.tex
\subsection{PoLaRIS-InvDes}
\label{sec:PoLaRIS-invdes}

\newcommand{\namemaps}{\texttt{MAPS}\xspace}
\newcommand{\namemapsdata}{\texttt{MAPS-Data}\xspace}
\newcommand{\namemapstrain}{\texttt{MAPS-Train}\xspace}
\newcommand{\namemapsinvdes}{\texttt{MAPS-InvDes}\xspace}
\newcommand{\nameboson}{\texttt{BOSON}$^{-1}$\xspace}

PoLaRIS-InvDes consists of two key components, \nameboson\cite{boson} and \namemaps\cite{maps}, to enable robust and efficient adjoint-based photonic device inverse design. 
Specifically, \nameboson formulates the inverse design task as a stochastic, fabrication-aware subspace optimization problem that explicitly models fabrication and operational variations, mitigates optimization difficulty via gradient-enhanced landscape reshaping, and achieves robust convergence through adaptive variation-aware sampling. 
\namemaps, on the other hand, is a modular AI-augmented inverse design infrastructure that provides a unified platform for dataset generation, model training, and design optimization. 
It includes tools for generating multi-fidelity, richly labeled photonic device data, supports training of AI models with physics- and data-driven objectives, and integrates seamlessly with an adjoint-based inverse design loop to accelerate optimization. 
Together, these two components form a comprehensive and extensible framework for advancing scalable, variation-robust photonic inverse design.
\input{figtex/fig_boson}
In \nameboson, as shown in Fig.~\ref{fig:boson}, we attribute the \textbf{key limitations of standard adjoint-based photonic inverse design} to three major factors: 
(1) \textbf{Sparse optimization objectives}, typically defined only at the output port, lead to poorly structured loss landscapes with sharp, undesired local optima and weak gradient signals; 
(2) \textbf{Fabrication and variation models} are often oversimplified, failing to capture the true impact of lithography, etching, and operational perturbations, which results in severe post-fabrication performance degradation; and 
(3) \textbf{Exhaustive variation sampling} methods, such as corner sweeping or Monte Carlo sampling, impose prohibitive simulation costs and dilute optimization focus on likely scenarios. 
To tackle these challenges, \nameboson formulates inverse design as a fabrication-aware, stochastic subspace optimization problem. 
It explicitly embeds \textbf{differentiable lithography and etching models into the optimization} loop to constrain the search within the manufacturable subspace. 
To mitigate the effects of sparse objectives and poor gradient flow, it introduces \textbf{auxiliary constraints that reshape the loss landscape and provide dense supervision}. 
Moreover, a \textbf{conditional subspace relaxation mechanism} enables the optimizer to escape suboptimal traps by constructing high-dimensional tunnels between the fabricable and idealized design spaces. 
Finally, \nameboson adopts an adaptive sampling strategy that combines axial and worst-case sampling to efficiently capture key variation modes while reducing the number of required simulations from exponential to linear scale. 
Together, these techniques enable \nameboson to produce compact, high-performance designs that are robust to real-world fabrication and operational uncertainties.

While \nameboson improves the robustness and sampling efficiency of adjoint-based photonic inverse design through stochastic formulation, subspace regularization, and adaptive variation-aware optimization, it fundamentally relies on iterative numerical solutions of Maxwell’s equations to compute gradients, resulting in high computational cost and limited scalability. 
To further accelerate the inverse design process and enable large-scale deployment, recent research has explored the integration of \emph{AI-based surrogate models to replace conventional solvers}. 
However, this introduces new methodological \underline{challenges}: (1) the absence of standardized datasets; (2) the lack of reproducible evaluation metrics for comparing AI models; and (3) the often overstated effectiveness of existing surrogates, which, while visually plausible in forward predictions, frequently \emph{fail to provide reliable gradients} necessary for optimization, thereby limiting their practical applicability in replacing numerical solvers.
To address these challenges, we \textbf{propose \namemaps, a modular, open-source infrastructure for AI-augmented photonic simulation and inverse design}, as shown in Fig.~\ref{fig:maps}. 
It unifies data generation, model training, and fabrication-aware optimization into a cohesive framework. 
Specifically, \namemapsdata provides a flexible data acquisition engine capable of generating multi-fidelity, richly annotated datasets with optimization-aware sampling strategies. 
\namemapstrain supports customizable training pipelines for data-driven and physics-informed models, incorporating standardized evaluation metrics such as field prediction accuracy and adjoint gradient alignment. 
\namemapsinvdes abstracts the complexity of adjoint-based optimization while enabling seamless integration with pretrained neural solvers and differentiable fabrication models. 
In conjunction with \nameboson, \namemaps facilitates scalable and variation-robust photonic inverse design, establishing a foundation for reproducible AI-for-optics research and real-world deployment.
\input{figtex/fig_maps}

%% file: figtex/fig_boson.tex
\begin{figure}
    \centering
    \includegraphics[width=0.7\columnwidth]{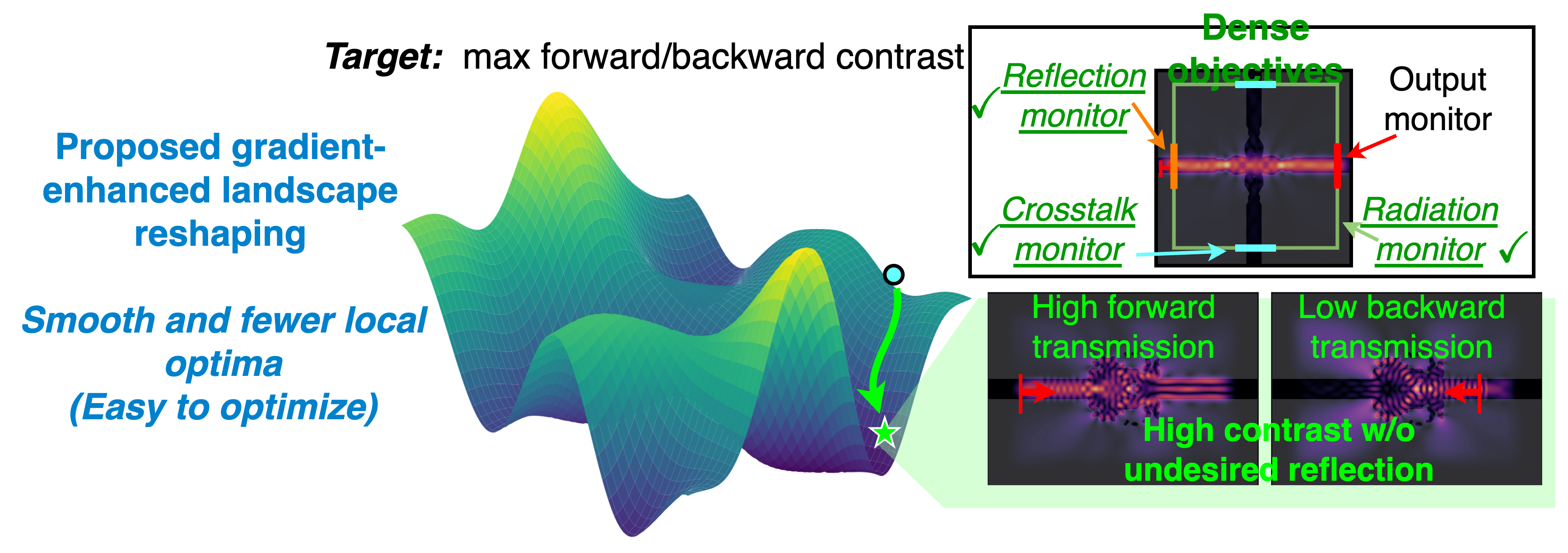}
    \caption{Advanced optimization algorithms ensure better optimality and convergence in photonic device inverse design from \nameboson~\cite{boson}.}
    \label{fig:boson}
     \vspace{-5pt}
\end{figure}

%% file: figtex/fig_maps.tex
\begin{figure}
    \centering
    \includegraphics[width=\columnwidth]{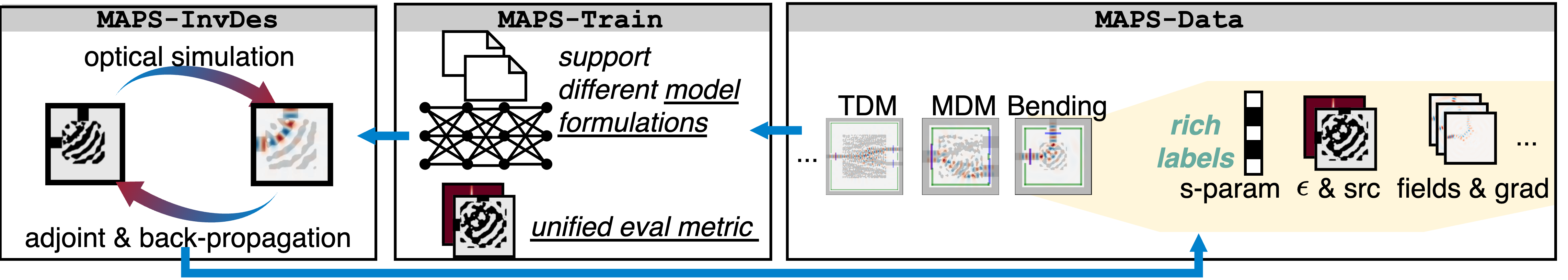}
    \caption{Illustration of AI-assisted photonic device simulation and inverse design infrastructure \namemaps~\cite{maps}.}
    \label{fig:maps}
     \vspace{5pt}
\end{figure}

%% file: doc/3_2_PhyDesign.tex
\subsection{PoLaRIS-PnR}
\label{sec:PoLaRIS-PnR}
\newcommand{\nameplacer}{\texttt{Apollo}\xspace}
\newcommand{\namerouter}{\texttt{LiDAR}\xspace}

PoLaRIS-PnR is formed by integrating a placer \textbf{\nameplacer}~\cite{PLACE_ICCAD2025_Zhou} and a router \textbf{\namerouter}~\cite{LiDAR_ISPD_Zhou, LiDAR2_ARXIV_Zhou} to achieve end-to-end layout generation for PICs. 
\nameplacer is a GPU-accelerated, waveguide-routing-informed analytical placer specifically tailored for PICs. 
It generates highly routable photonic placement solutions by considering photonic-specific constraints during placement optimization, while \namerouter is a curvy-aware waveguide detailed router based on A$^\ast$ search. 
It generates waveguide routes that respect the minimum bending radius constraints and supports dynamic waveguide crossing insertion. 
This eliminates the need for manual planning and generates low insertion loss routing results.

\input{figtex/fig_place_challenge}
\noindent\textbf{PIC Placement} -- 
In PIC placement, \textbf{routing is a dominant constraint}. 
Unlike VLSI, where routability can often be improved via cell inflation or whitespace, PICs face unique challenges as shown in Fig.~\ref {fig:place_challenge}. 
Waveguides are orientation-sensitive, curvy, and constrained to very few routing layers. 
Bends and crossings are area-intensive, and improper port orientations can cause excessive detours, leading to unroutable or lossy connections.
Furthermore, photonic devices often have high port density. 
For example, a multi-mode interference (MMI) device may contain many ports tightly packed along its edges, requiring additional whitespace for waveguide escapes and crossings. 
To address these challenges, \nameplacer introduces \underline{several routing-aware innovations}. 
\ding{202}~To account for \emph{port orientation} and reduce routing overhead from curvy bends, \namerouter introduces an \textbf{asymmetric bending-aware wirelength model}, called the $\textrm{cosWA}$ function. 
\ding{203}~To estimate spacing demands from bends, crossings, and multiport access, a routing-informed net spacing model is used, incorporating both \textbf{port density}, the clustering of co-directional ports, and estimated \textbf{routing congestion} to reflect potential waveguide crossings.

Beyond routability, functional and robust PIC layouts must satisfy \textbf{domain-specific constraints} reflecting physical and performance requirements. 
For example, thermo-optic phase shifters require spacing to mitigate thermal crosstalk, and certain components or nets demand matched routing topologies for phase coherence in interferometric circuits.
\ding{204}~\nameplacer addresses these design constraints via: (1) \textbf{cell inflation}, which reserves space for thermally sensitive components, and (2) \textbf{projected gradient descent}, which promotes alignment and topological uniformity as shown in Fig.~\ref{fig:place_constraints}. 
\input{figtex/fig_place_constraints}

From an optimization standpoint, the extreme size and aspect ratio diversity of photonic components, ranging from $2 \times 2~\mu\mathrm{m}^2$ splitters to millimeter-scale modulators, poses challenges similar to mixed-size VLSI placement, which is combinatorial and ill-conditioned.
To address this, \nameplacer employs a \textbf{Blockwise Adaptive Barzilai-Borwein (BBB)} step size within a Nesterov-accelerated gradient descent framework. 
This approach decouples the updates of large and small components, stabilizing large cell movement while preserving fine-grained optimization for smaller cells, leading to a more scalable and robust placement process.

\noindent\textbf{PIC Routing} --
After placement, \textbf{\namerouter} performs waveguide routing and distinguishes itself from prior works by generating \textbf{DRV-free, real GDS-II layouts}. 
A key constraint is the bending radius, typically 5--10~$\mu$m, as sharp turns are forbidden in PICs. 
While some works~\cite{NP_iccad2013_proton} ignore this during path search and apply smoothing post hoc, such strategies often fail under tight layout and routing constraints.
To address this, \namerouter introduces a \textbf{curvy-aware A$^\ast$ search}, which considers the node orientation and bending radius to generate the neighbors. 
An \textbf{orientation-aware bitmap} is proposed to track occupied regions and enforce spacing rules, ensuring legal paths and supporting \textbf{dynamic waveguide crossing insertion} when two waveguides are orthogonal, type-compatible, and sufficiently spaced, eliminating the need for manual crossing planning.

\textbf{Routability remains a major challenge} due to limited routing layers, strict crossing conditions, and precise alignment required by directional waveguide ports. 
To ensure robust access, \namerouter introduces \underline{four techniques}: 
(1) \emph{Port Propagation} extends ports to the chip boundary along their orientation; 
(2) \emph{Bending-Aware Region Reservation} reserves nearby grid cells to prevent interference; 
(3) \emph{Congested-Port Spreading} offsets dense port clusters to avoid overlap; and 
(4) \emph{Staggered Channel Planning} introduces access-point offsets to reduce contention. 
To further mitigate \texttt{routing conflicts}, \namerouter uses a \textbf{group-based net ordering} strategy that clusters same-orientation ports and reserves resources via a \textbf{group congestion penalty}. 
A \textbf{local rip-up and reroute} refinement further removes unnecessary crossings, reducing loss and improving layout quality.
In addition, PIC detailed routing is \texttt{time-consuming} due to long waveguides and complex geometric constraints. 
To improve scalability, we extend \namerouter with a \textbf{hierarchical routing framework} that promotes sub-circuit reuse and reduces inter-module conflicts~\cite{LiDAR2_ARXIV_Zhou}. 
Additionally, \textbf{offset neighbors} are introduced to expand routing flexibility while preserving feasibility. 
Together, these enhancements significantly improve routing efficiency and layout quality in large-scale PICs.

\input{figtex/fig_pnr_results}
By integrating \nameplacer and \namerouter, PoLaRIS-PnR enables the first open-source PIC layout generation toolflow. 
Figure~\ref{fig:pnr_result} shows the PIC layout generated by the PoLaRIS-PnR. 
\nameplacer ensures routable and constraint-aware placement, while \namerouter delivers a DRV-free waveguide routing solution, providing a practical and scalable solution for large-scale PIC design automation.

%% file: figtex/fig_place_challenge.tex
\begin{figure}
    \centering
    \includegraphics[width=\columnwidth]{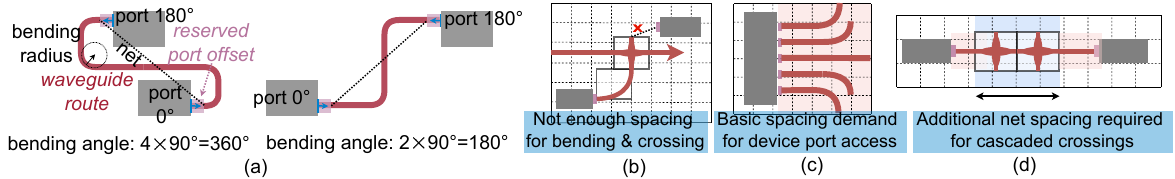}
    \caption{Challenges of the automated PIC placement~\cite{PLACE_ICCAD2025_Zhou}.}
    \label{fig:place_challenge}
     \vspace{-5pt}
\end{figure}

%% file: figtex/fig_place_constraints.tex
\begin{figure}
    \centering
    \includegraphics[width=\columnwidth]{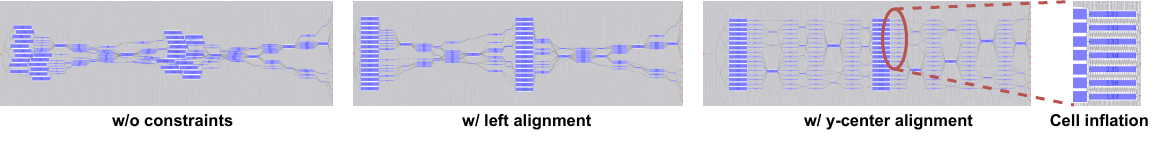}
    \vspace{-10pt}
    \caption{Proper placement constraints ensure high PIC layout quality~\cite{PLACE_ICCAD2025_Zhou}.}
    \label{fig:place_constraints}
     \vspace{-10pt}
\end{figure}

%% file: figtex/fig_pnr_results.tex
\begin{figure}
    \centering
    \includegraphics[width=\columnwidth]{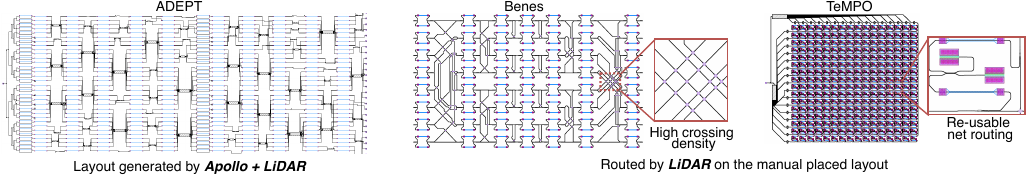}
    \vspace{-10pt}
    \caption{PIC layout visualization from \nameplacer~\cite{PLACE_ICCAD2025_Zhou} and \namerouter~\cite{LiDAR2_ARXIV_Zhou}.}
    \label{fig:pnr_result}
     \vspace{-5pt}
\end{figure}

%% file: doc/4_Direction.tex
\section{Conclusion and Future Direction}
\label{sec:direction}

In this work, we present \textbf{PoLaRIS}, a comprehensive intelligent design automation framework to address the critical bottlenecks in large-scale PIC design. 
By unifying a fabrication-aware inverse design engine with a routing-informed automated layout generator, PoLaRIS significantly accelerates the design cycle, enhances performance, and lowers the barrier to entry for this transformative technology. 
The successful generation of complex, DRV-free layouts demonstrates a pivotal step toward robust and scalable PIC design automation.

Looking ahead, the automation of electronic-photonic design is a cornerstone for unlocking the next generation of computing, communication, and sensing systems. The continued evolution of this field promises to enable groundbreaking hardware platforms, from ultra-high-bandwidth optical interconnects for data centers to energy-efficient AI accelerators and novel bio-photonic sensors. 
In the future, embedding differentiable photonic simulators directly into the EPDA loop will enable on-the-fly trade-offs between insertion loss, crosstalk, and system footprint. 
Expanding beyond single‑layer waveguides to multi‑layer and 3D interconnects will unlock unprecedented integration density. Additionally, leveraging machine learning-guided heuristics and incorporating advanced yield and variability models will ensure robust, first-pass success in fabrication and reduce late-stage design iterations. 
Finally, developing a true photonic-electronic co-design framework will be critical to holistically optimize these complex systems.
By creating a more intelligent and accessible design ecosystem, we envision a future where PIC technology can be rapidly harnessed to solve some of the world's most pressing technological challenges, paving the way for continued innovation in integrated photonics.

%% file: main.bbl
\begin{thebibliography}{10}

\bibitem{NP_Nature_ahmed}
Ahmed, S.~R., Baghdadi, R., Bernadskiy, M., Bowman, N., Braid, R., Carr, J., Chen, C., Ciccarella, P., Cole, M., Cooke, J., et~al., ``Universal photonic artificial intelligence acceleration,'' {\em Nature}~{\bf 640}(8058),  368--374 (2025).

\bibitem{NP_Light_Zhou}
Zhou, H., Dong, J., Cheng, J., Dong, W., Huang, C., Shen, Y., Zhang, Q., Gu, M., Qian, C., Chen, H., et~al., ``Photonic matrix multiplication lights up photonic accelerator and beyond,'' {\em Light: Science \& Applications}~{\bf 11}(1),  30 (2022).

\bibitem{NP_DATE2020_popstar}
Thonnart, Y., Bernab{\'e}, S., Charbonnier, J., Bernard, C., Coriat, D., Fuguet, C., Tissier, P., Charbonnier, B., Malhouitre, S., Saint-Patrice, D., et~al., ``Popstar: A robust modular optical noc architecture for chiplet-based 3d integrated systems,'' in [{\em DATE 2020-Design, Automation \& Test in Europe Conference \& Exhibition}{\nolinebreak\hspace{0.1em}]},   1456--1461, IEEE (2020).

\bibitem{NP_book_chrostowski}
Chrostowski, L. and Hochberg, M.,  [{\em Silicon photonics design: from devices to systems}{\nolinebreak\hspace{0.1em}]}, Cambridge University Press (2015).

\bibitem{NP_minkov2020inverse}
Minkov, M., Williamson, I.~A., Andreani, L.~C., Gerace, D., Lou, B., Song, A.~Y., Hughes, T.~W., and Fan, S., ``Inverse design of photonic crystals through automatic differentiation,'' {\em Acs Photonics}~{\bf 7}(7),  1729--1741 (2020).

\bibitem{NP_khoram2020controlling}
Khoram, E., Qian, X., Yuan, M., and Yu, Z., ``Controlling the minimal feature sizes in adjoint optimization of nanophotonic devices using b-spline surfaces,'' {\em Optics Express}~{\bf 28}(5),  7060--7069 (2020).

\bibitem{NP_li2020fourier}
Li, Z., Kovachki, N., Azizzadenesheli, K., Liu, B., Bhattacharya, K., Stuart, A., and Anandkumar, A., ``Fourier neural operator for parametric partial differential equations,'' {\em arXiv preprint arXiv:2010.08895}  (2020).

\bibitem{NP_tran2021factorized}
Tran, A., Mathews, A., Xie, L., and Ong, C.~S., ``Factorized fourier neural operators,'' {\em arXiv preprint arXiv:2111.13802}  (2021).

\bibitem{NP_zhang2024sinenet}
Zhang, X., Helwig, J., Lin, Y., Xie, Y., Fu, C., Wojtowytsch, S., and Ji, S., ``Sinenet: Learning temporal dynamics in time-dependent partial differential equations,'' {\em arXiv preprint arXiv:2403.19507}  (2024).

\bibitem{gu2022NeurOLight}
Gu, J., Gao, Z., Feng, C., Zhu, H., Chen, R.~T., Boning, D.~S., and Pan, D.~Z., ``Neurolight: A physics-agnostic neural operator enabling parametric photonic device simulation,'' in [{\em Conference on Neural Information Processing Systems (NeurIPS)}{\nolinebreak\hspace{0.1em}]},  (2022).

\bibitem{zhu2024Pace}
Zhu, H., Cong, W., Chen, G., Ning, S., Chen, R., Gu, J., and Pan, D.~Z., ``Pace: Pacing operator learning to accurate optical field simulation for complicated photonic devices,'' in [{\em Conference on Neural Information Processing Systems (NeurIPS)}{\nolinebreak\hspace{0.1em}]},  (2024).

\bibitem{kossaifi2024neural}
Kossaifi, J., Kovachki, N., Li, Z., Pitt, D., Liu-Schiaffini, M., George, R.~J., Bonev, B., Azizzadenesheli, K., Berner, J., and Anandkumar, A., ``A library for learning neural operators,'' (2024).

\bibitem{kovachki2021neural}
Kovachki, N.~B., Li, Z., Liu, B., Azizzadenesheli, K., Bhattacharya, K., Stuart, A.~M., and Anandkumar, A., ``Neural operator: Learning maps between function spaces,'' {\em CoRR}~{\bf abs/2108.08481} (2021).

\bibitem{zhang2024sinenet}
Zhang, X., Helwig, J., Lin, Y., Xie, Y., Fu, C., Wojtowytsch, S., and Ji, S., ``Sinenet: Learning temporal dynamics in time-dependent partial differential equations,'' in [{\em The Twelfth International Conference on Learning Representations}{\nolinebreak\hspace{0.1em}]},  (2024).

\bibitem{pcma2024pic2osim}
Ma, P., Yang, H., Gao, Z., Boning, D.~S., and Gu, J., ``{PIC$^2$O-Sim: A physics-inspired causality-aware dynamic convolutional neural operator for ultra-fast photonic device time-domain simulation},'' {\em APL Photonics}~{\bf 10},  036104 (03 2025).

\bibitem{NP_chen2020design}
Chen, M., Jiang, J., and Fan, J.~A., ``Design space reparameterization enforces hard geometric constraints in inverse-designed nanophotonic devices,'' {\em ACS Photonics}~{\bf 7}(11),  3141--3151 (2020).

\bibitem{NP_gershnabel2022reparameterization}
Gershnabel, E., Chen, M., Mao, C., Wang, E.~W., Lalanne, P., and Fan, J.~A., ``Reparameterization approach to gradient-based inverse design of three-dimensional nanophotonic devices,'' {\em ACS Photonics}~{\bf 10}(4),  815--823 (2022).

\bibitem{hammond2022high}
Hammond, A.~M., Oskooi, A., Chen, M., Lin, Z., Johnson, S.~G., and Ralph, S.~E., ``High-performance hybrid time/frequency-domain topology optimization for large-scale photonics inverse design,'' {\em Optics Express}~{\bf 30}(3),  4467--4491 (2022).

\bibitem{wang2019robust}
Wang, E.~W., Sell, D., Phan, T., and Fan, J.~A., ``Robust design of topology-optimized metasurfaces,'' {\em Optical Materials Express}~{\bf 9}(2),  469--482 (2019).

\bibitem{wang2011robust}
Wang, F., Jensen, J.~S., and Sigmund, O., ``Robust topology optimization of photonic crystal waveguides with tailored dispersion properties,'' {\em Journal of the Optical Society of America B}~{\bf 28}(3),  387--397 (2011).

\bibitem{schevenels2011robust}
Schevenels, M., Lazarov, B.~S., and Sigmund, O., ``Robust topology optimization accounting for spatially varying manufacturing errors,'' {\em Computer Methods in Applied Mechanics and Engineering}~{\bf 200}(49-52),  3613--3627 (2011).

\bibitem{gershnabel2022reparameterization}
Gershnabel, E., Chen, M., Mao, C., Wang, E.~W., Lalanne, P., and Fan, J.~A., ``Reparameterization approach to gradient-based inverse design of three-dimensional nanophotonic devices,'' {\em ACS Photonics}~{\bf 10}(4),  815--823 (2022).

\bibitem{hammond2021photonic}
Hammond, A.~M., Oskooi, A., Johnson, S.~G., and Ralph, S.~E., ``Photonic topology optimization with semiconductor-foundry design-rule constraints,'' {\em Optics Express}~{\bf 29}(15),  23916--23938 (2021).

\bibitem{chen2020design}
Chen, M., Jiang, J., and Fan, J.~A., ``Design space reparameterization enforces hard geometric constraints in inverse-designed nanophotonic devices,'' {\em ACS Photonics}~{\bf 7}(11),  3141--3151 (2020).

\bibitem{mao2023multi}
Mao, S., Cheng, L., Chen, H., Liu, X., Geng, Z., Li, Q., and Fu, H., ``Multi-task topology optimization of photonic devices in low-dimensional fourier domain via deep learning,'' {\em Nanophotonics}~{\bf 12}(5),  1007--1018 (2023).

\bibitem{chrostowski2016schematic}
Chrostowski, L., Lu, Z., Fl{\"u}ckiger, J., Pond, J., Klein, J., Wang, X., Li, S., Tai, W., Hsu, E.~Y., Kim, C., et~al., ``Schematic driven silicon photonics design,'' in [{\em Smart Photonic and Optoelectronic Integrated Circuits XVIII}{\nolinebreak\hspace{0.1em}]},   {\bf 9751},  9--22, SPIE (2016).

\bibitem{GDSFactory}
Matres, J. et~al., ``Gdsfactory.'' \url{https://github.com/gdsfactory/gdsfactory} (2024).

\bibitem{NP_DAC2018_planaronoc}
Chuang, Y.-K., Chen, K.-J., Lin, K.-L., Fang, S.-Y., Li, B., and Schlichtmann, U., ``Planaronoc: concurrent placement and routing considering crossing minimization for optical networks-on-chip,'' in [{\em Proceedings of the 55th Annual Design Automation Conference}{\nolinebreak\hspace{0.1em}]},   1--6 (2018).

\bibitem{NP_DAC2009_ding}
Ding, D., Zhang, Y., Huang, H., Chen, R.~T., and Pan, D.~Z., ``O-router: an optical routing framework for low power on-chip silicon nano-photonic integration,'' in [{\em Proceedings of the 46th annual design automation conference}{\nolinebreak\hspace{0.1em}]},   264--269 (2009).

\bibitem{NP_ISPD2019_psion}
Truppel, A., Tseng, T.-M., Bertozzi, D., Alves, J.~C., and Schlichtmann, U., ``Psion: Combining logical topology and physical layout optimization for wavelength-routed onocs,'' in [{\em Proceedings of the 2019 International Symposium on Physical Design}{\nolinebreak\hspace{0.1em}]},   49--56 (2019).

\bibitem{boson}
Ma, P., Gao, Z., Begovic, A., Zhang, M., Yang, H., Ren, H., Huang, R., Boning, D.~S., and Gu, J., ``{BOSON$^{-1}$: Understanding and Enabling Physically-Robust Photonic Inverse Design with Adaptive Variation-Aware Subspace Optimization},'' in [{\em 2025 Design, Automation \& Test in Europe Conference (DATE)}{\nolinebreak\hspace{0.1em}]},  (2025).

\bibitem{maps}
Mal, P., Gao, Z., Zhang, M., Yang, H., Ren, M., Huang, R., Boning, D.~S., and Gu, J., ``{MAPS: Multi-Fidelity AI-Augmented Photonic Simulation and Inverse Design Infrastructure},'' in [{\em 2025 Design, Automation \& Test in Europe Conference (DATE)}{\nolinebreak\hspace{0.1em}]},   1--6 (2025).

\bibitem{PLACE_ICCAD2025_Zhou}
Zhou, H., Yang, H., Nicholas, G., Ren, H., Rena, H., and Gu, J., ``Apollo: Automated routing-informed placement for large-scale photonic integrated circuits,'' in [{\em International Conference on Computer-Aided Design (ICCAD)}{\nolinebreak\hspace{0.1em}]},  (2025).

\bibitem{LiDAR_ISPD_Zhou}
Zhou, H., Zhu, K., and Gu, J., ``{LiDAR: Automated Curvy Waveguide Detailed Routing for Large-Scale Photonic Integrated Circuits},'' in [{\em Proceedings of the 2025 International Symposium on Physical Design}{\nolinebreak\hspace{0.1em}]},   64--72 (2025).

\bibitem{LiDAR2_ARXIV_Zhou}
Zhou, H., Yang, H., Ying, Z., Gangi, N., Ren, H., Matres, J., Gu, J., et~al., ``{LiDAR 2.0: Hierarchical Curvy Waveguide Detailed Routing for Large-Scale Photonic Integrated Circuits},'' {\em arXiv preprint arXiv:2505.17239}  (2025).

\bibitem{NP_iccad2013_proton}
Boos, A., Ramini, L., Schlichtmann, U., and Bertozzi, D., ``Proton: An automatic place-and-route tool for optical networks-on-chip,'' in [{\em 2013 IEEE/ACM International Conference on Computer-Aided Design (ICCAD)}{\nolinebreak\hspace{0.1em}]},   138--145, IEEE (2013).

\end{thebibliography}
